\documentclass[prb,twocolumn,showpacs]{revtex4-1}
\usepackage[latin1]{inputenc}
\usepackage{graphicx}
\usepackage{amssymb}
\usepackage{amsmath}
\usepackage{xspace}
\usepackage{dcolumn}
\usepackage{bm}
\usepackage{multirow}
\usepackage{color}
\usepackage[extra]{tipa}

\newfont{\tensy}{cmsy10}

\newcommand{\ie}[0]{{\it i.e.\@\xspace}}

\newcommand{\etal}[0]{{\it et al.\@\xspace}}

\newcommand{\rmi}{\text{i}}
\newcommand{\UP}[0]{\uparrow}
\newcommand{\DO}[0]{\downarrow}

\newcommand{\on}{\hat{n}}

\newcommand{\oq}{\mbox{$\frac{1}{4}$}}

\newcommand{\om}[0]{\omega}
\newcommand{\Ep}{\varepsilon_\text{p}}

\newcommand{\kF}{k_\text{F}}

\newcommand{\nag}{{\phantom{\dag}}}

\newcommand{\las}[0]{\langle}
\newcommand{\ras}[0]{\rangle}

\newcommand{\ra}[0]{\right\ras}
\newcommand{\ket}[1]{\left|#1\ra}

\begin{document}

%%%%%%%%%%%%%%%%%%%%%%%%%%%%%%%%%%%%%%%%%%%%%%%%%%%%%%%%%%%%%%%%%%%%%%%%%%%%%
%%%%%%%%%%%%%%%%%%%%% TITLE & ABSTRACT %%%%%%%%%%%%%%%%%%%%%%%
%%%%%%%%%%%%%%%%%%%%%%%%%%%%%%%%%%%%%%%%%%%%%%%%%%%%%%%%%%%%%%%%%%%%%%%%%%%%%

\title{Charge and spin correlations of a Peierls insulator after a quench}

\author{Martin Hohenadler}
\affiliation{\mbox{Institut f\"ur Theoretische Physik und Astrophysik,
    Universit\"at W\"urzburg, 97074 W\"urzburg, Germany}}

\begin{abstract}
Electron-phonon coupling plays a central role for time-dependent
phenomena in condensed matter, for example in photo-excitation experiments.
We use the continuous-time quantum Monte Carlo method to study the real-time
evolution of charge and spin correlation functions of a Peierls insulator
after a quench to a noninteracting Hamiltonian. This approach gives exact
results, and fully takes into account quantum phonon effects without relying
on a Hilbert space truncation. It is also free from a dynamical sign
problem. The observed time dependence is compared to free-fermion time
evolution starting from a dimerized state. Our exact results provide a
benchmark for more realistic calculations, and may be directly applicable to
experiments with cold atoms or trapped ions.
\end{abstract}

\date{\today}


\maketitle

\section{Introduction}\label{sec:intro}

The enormous current interest in the time evolution or nonequilibrium
behavior of quantum systems is driven by significant developments in the
field of experimental physics. Of particular importance are time-resolved
studies of cold atoms in optical lattices following the seminal work by
Greiner \etal,\cite{Greiner00} and the progress with femtosecond
spectroscopy,\cite{PhysRevLett.105.067001,PhysRevLett.105.257001,PhysRevLett.105.246402,Wall2011}
including time-resolved photoemission
spectroscopy.\cite{PhysRevLett.99.197001,PhysRevLett.107.097002} 
These methods enable researchers to carry out pump-probe experiments and to study photo-excitation and
photo-induced phase transitions.\cite{Chollet07012005,PhysRevLett.98.097402,PhysRevLett.99.116401,PhysRevLett.102.066404,Hellmann2010,PhysRevB.81.073102,JPSJ.79.034708,PhysRevLett.105.246402}

Theoretical studies are particularly difficult for systems with
strong correlations, as relevant for several classes of
materials. Whereas much progress has been made in developing methods 
to study correlated systems in equilibrium, the description of time-dependent
or nonequilibrium problems is extremely challenging. As a result, it is
currently not possible to reliably describe experiments. Promising directions
of research include time-dependent density matrix renormalization group
methods,\cite{Schollwock201196} extensions of quantum cluster methods such as
dynamical mean-field theory\cite{SchmidtMonien02,PhysRevLett.97.266408} and
the variational cluster approach,\cite{PhysRevB.84.115145} as well as
extensions of Luttinger liquid theory.\cite{PhysRevLett.109.126406} A review
of theoretical methods for correlated systems has been given by Eckstein
\etal\cite{Ecksteinreview} An important open question for gapless one-dimensional
systems is if the time dependence of generic lattice models is indeed fully
captured by Luttinger liquid theory.\cite{Becca2013}

Whereas these ideas have successfully been applied to models of interacting
spins, fermions or bosons, much less is known about phenomena related to
electron-phonon interaction. The latter plays a crucial role since phonons
are often the fastest channel available for thermalization via
scattering.\cite{arXiv:1212.4841} However, the infinite bosonic Hilbert space
as well as the resulting retarded electron-electron interaction pose a
serious challenge to theoretical methods. Whereas exact results have been
obtained for a single
polaron,\cite{PhysRevLett.109.236402,PhysRevB.75.014307,PhysRevB.83.075104} a
single bipolaron,\cite{PhysRevB.85.144304} and a single-molecule
device,\cite{Vinkler2011} lattice models with finite band fillings are much
more challenging to study. Recent work along these lines makes use of the density
matrix renormalization group on small systems,\cite{JPSJ.81.013701} the
so-called double-phonon cloud method,\cite{PhysRevLett.109.176402} restricted
phonon Hilbert spaces,\cite{PhysRevB.84.195109} as well as classical
phonons.\cite{PhysRevB.79.125118}

A particularly interesting class of experiments involves photo-induced phase
transitions across a metal-insulator Peierls
transition;\cite{Chollet07012005,PhysRevLett.98.097402,PhysRevLett.99.116401,Schmitt2008,PhysRevLett.102.066404,Hellmann2010,PhysRevB.81.073102,JPSJ.79.034708,PhysRevLett.105.246402}
see Ref.~\onlinecite{Yonemitsu20081} for a review. 
A laser pulse excites the system out of the insulating, charge-ordered state,
as visible for example from the band structure (probed by time and angle-resolved photoemission
spectroscopy)\cite{PhysRevB.81.073102} or from the
reflectivity.\cite{PhysRevLett.102.066404} After the photo-induced melting of
charge order, the system relaxes back towards its original state. Such
experiments also reveal the typically very different time scales for electron
and ion dynamics. So far, a photo-induced phase transition of a Peierls
system has been considered theoretically in
Ref.~\onlinecite{PhysRevB.84.195109} using a classical phonon
approximation. Since nonlocal correlations are a defining feature of charge
ordered states, local approximations such as dynamical-mean field theory are
of limited use.

Motivated by the ongoing effort to improve the available theoretical tools
for the study of electron-phonon systems, we consider here a  quench
from a Peierls insulator to noninteracting fermions by switching off the
electron-phonon interaction. The model we use is the one-dimensional Holstein
model.  The time evolution of charge and spin correlation functions after the
quench is studied with an extension of the continuous-time quantum Monte
Carlo method.\cite{Rubtsov05} Although such a quench does not fully capture the
complexity encountered in pump-probe experiments, it
provides us with a rare opportunity to obtain exact results for the time
evolution of an electron-phonon system. Our findings can serve as a
nontrivial benchmark for future numerical calculations (for example with
methods that rely on a truncation of the phonon Hilbert space), and may even
be relevant for experimental realizations of electron-phonon models in
optical
lattices\cite{PhysRevA.84.051401,1367-2630-14-3-033019,PhysRevA.72.033609,BlochSSH}
or ion traps.\cite{PhysRevLett.109.200501,PhysRevLett.109.250501}

The paper is organized as follows. In Sec.~\ref{sec:model}, we discuss the
model, in Sec.~\ref{sec:method}, we provide details about the method used, in
Sec.~\ref{sec:results}, we present our results, and Sec.~\ref{sec:conclusions}
contains our conclusions.

\section{Model}\label{sec:model}

The Peierls metal-insulator transition in one dimension can be investigated
in the framework of Holstein's molecular crystal model,\cite{Ho59a} defined by the Hamiltonian
\begin{align}\label{eq:model}
  \hat{H}  
  &=
  \sum_{k\sigma}\epsilon(k)c^\dag_{k\sigma} c^\nag_{k\sigma}
  + \sum_{i} \left( \frac{\hat{P}_{i}^2}{2M} + \frac{K \hat{Q}_{i}^2}{2}
     \right)
  %t\sum_{i \sigma} \left( c^{\dag}_{i\sigma} c^\nag_{i+1\sigma} + \text{H.c.} \right)
  \\\nonumber
  &\quad
  - g(t) \sum_{i} \hat{Q}_{i} \left( \hat{n}_{i}-1\right) 
  \,.
\end{align}
The first term, referred to as $\hat{H}_0$ in the following, describes
nearest-neighbor hopping of electrons with the usual tight-binding band
structure $\epsilon(k)=-2 J\cos k$; we have set the lattice constant equal
to one, and take the hopping integral as the energy unit. The second and
third terms correspond to the lattice degrees of freedom and the
electron-phonon coupling. The lattice is described by means of harmonic
oscillators with frequency $\om_0=\sqrt{K/M}$, displacement $\hat{Q}_i$ and
momentum $\hat{P}_i$, and the coupling is of the density-displacement type
proposed by Holstein;\cite{Ho59a} $g(t)$ is the coupling constant.  We use
the usual definitions $\hat{n}_{i\sigma} = c^{\dag}_{i\sigma }
c^\nag_{i\sigma }$, $\on_i = \sum_\sigma \on_{i\sigma}$, and $n=\las
\on_i\ras$.  A useful dimensionless coupling parameter is given by the ratio
$\lambda=2\Ep/W$, where $\Ep=g^2/2K $ is the polaron binding energy, and
$W=4J$ is the bare bandwidth. We consider a half filled band, \ie, $\las
\on_i \ras =1$.

As discussed in Sec.~\ref{sec:intro}, we study a quantum quench 
to a noninteracting Hamiltonian. This is achieved by
a time-dependent coupling constant $g(t)$. Explicitly, we take
\begin{equation}\label{eq:quench}
  g(t) 
  =
  \begin{cases}
    g\,, & t= 0\,,\\
    0\,, & t> 0\,.
  \end{cases}    
\end{equation}
Because there is no coupling between electrons and the lattice after such a
quench, the time evolution will be determined by the hopping term $\hat{H}_0$
only.

\section{Method}\label{sec:method}

The continuous-time quantum Monte Carlo (CTQMC) method is based on a
weak-coupling Dyson expansion of the path integral for the partition
function.\cite{Rubtsov05} For finite systems, all possible vertex
contributions can be summed stochastically, thereby giving exact results with
only statistical errors. The CTQMC method has been used quite extensively for
lattice and impurity problems, see Ref.~\onlinecite{Gull_rev} for a review.
Because it is formulated in terms of an action, the CTQMC method permits us
to simulate models with retarded interactions, which arise in the path
integral for electron-phonon models when integrating analytically over the lattice
degrees of freedom.\cite{Assaad07} This approach has been applied to study
a variety of electron-phonon models;
\cite{Hohenadler10a,Assaad08,HoAs12,PhysRevB.87.075149,Ho.As.Fe.12} numerical
details can be found in Ref.~\onlinecite{PhysRevB.87.075149}. Because the method relies
on an analytical integration over the phonon degrees of freedom, lattice
properties such as local displacements cannot be directly measured.

Quantum Monte Carlo simulations of time-dependent problems are inherently
limited by the dynamical sign problem related to the time evolution operator
$\exp(\rmi \hat{H} t)$. Progress has been made by considering simplified
situations where the sign problem can be avoided or limited, including quantum quenches to noninteracting
systems,\cite{PhysRevB.85.085129} and studies of (effective) impurity
problems.  For the CTQMC method, the first application to a
nonequilibrium problem involving phonons was achieved in
Ref.~\onlinecite{PhysRevLett.100.176403}, and a review of previous work is
found in Ref.~\onlinecite{Gull_rev}. Recently, the weak-coupling CTQMC method
was formulated on the Keldysh contour.\cite{PhysRevB.85.085129}

In principle, the method of Ref.~\onlinecite{PhysRevB.85.085129} can be
extended to a retarded electron-electron interaction. The resulting algorithm 
would, in principle, be suitable to study rather general
problems, but with the usual limitation to short times due to the
dynamical sign problem.\cite{PhysRevB.85.085129} Here, by applying ideas
from previous work using auxiliary-field methods,\cite{PhysRevB.85.085129}
we focus on quenches to noninteracting systems. In simulations of the time
evolution {\it   after} the electron-phonon coupling $g$ has been switched
off [see Eq.~(\ref{eq:quench})], interaction vertices are restricted to
imaginary times on the Keldysh contour, and there is no dynamical sign problem.

The numerical method proceeds as follows. The interacting Holstein
model~(\ref{eq:model}) is simulated with the CTQMC method using standard
Monte Carlo updates (addition and removal of vertices, flipping of Ising
spins\cite{Assaad07}) to obtain configurations reflecting the initial,
correlated state at time $t=0$.  The interacting single-particle Green's
function matrix at equal imaginary times,
\begin{equation}
  [\mathbf{G}_\sigma(t=0)]_{ij} = \las c^\nag_{i\sigma} c^\dag_{j\sigma}\ras\,,
\end{equation}
is calculated for a given vertex configuration from the Dyson equation.\cite{Rubtsov05,Assaad07}
From $\mathbf{G}_\sigma(t=0)$, all relevant correlation functions and
observables at $t=0$ can be obtained using Wick's theorem.\cite{Rubtsov05,Assaad07}
For quenches of the form~(\ref{eq:quench}), expectation values at times $t>0$
follow from the time-propagated Green's function
\begin{equation}\label{eq:propG}
  \mathbf{G}_\sigma(t) = e^{-\rmi \hat{H}_0 t}\, \mathbf{G}_\sigma(t=0)\,
  e^{\rmi \hat{H}_0 t}\,,
\end{equation}
in combination with Wick's theorem.  The additional numerical effort compared
to a time-independent simulation amounts to the evaluation of
Eq.~(\ref{eq:propG}) and of observables of interest for each value of $t$. Results for each
$t$ are averaged over different Monte Carlo configurations. Whereas
equal-time observables can easily be calculated in this way, time-displaced
correlation functions require a formulation on the full Keldsyh contour.

\section{Results}\label{sec:results}

For a half-filled band, the Holstein model~(\ref{eq:model}) describes a
transition from a metal to a Peierls insulator driven by the electron-phonon
interaction. At $T=0$, the Peierls state has long-rage charge order with
wavevector $q=2\kF$, and a gap in the single-particle
spectrum.\cite{Hirsch83a} In addition, a spin gap presumably exists for any
nonzero electron-phonon interaction, as a result of electrons pairing into
singlets.\cite{PhysRevB.87.075149} Whereas the model is dimerized and
insulating for any $\lambda>0$ in the mean-field limit
$\om_0=0$,\cite{Hirsch83a} numerical
results\cite{JeZhWh99,0295-5075-84-5-57001,ClHa05,hardikar:245103} indicate a
nonzero critical value for $\om_0>0$ as a result of quantum lattice
fluctuations. For an overview of previous work, see
Ref.~\onlinecite{PhysRevB.87.075149}.

Here, we consider parameters for which Hamiltonian~(\ref{eq:model}) is in the
insulating Peierls phase. Explicitly, we take $\lambda=0.5$, and two values
of the phonon frequency, $\om_0=0.5J$, and $5J$. According to quantum
Monte Carlo results,\cite{hardikar:245103} $\lambda=0.5$ is well inside the
ordered phase for $\om_0=0.5J$ (the critical coupling is about 0.25), and
just above the critical coupling for $\om_0=5J$. Since our simulations are
carried out at a low but finite temperature, the system will be in a thermal initial state
before the quench. However, the results shown below are practically identical
for $\beta J=L$ and $\beta J=2L$.

Among the equal-time observables accessible with the present method, the
real-space charge and spin correlation functions
\begin{align}\label{eq:correlators}\nonumber
  S_\rho(r)   &=  \las \on_r \on_0 \ras\,,\\
  S_\sigma(r)  &= \las \hat{S}^z_r \hat{S}^z_0 \ras\,,
\end{align}
are of particular interest. Since we consider a quench to the noninteracting
Hamiltonian $\hat{H}_0$, we compare the correlation functions to those of free
fermions, which have the form\cite{Schulz90}
\begin{align}\label{eq:noninteractingLL}
  S_\rho(r) = S_\sigma(r) = 
  -\frac{1}{(\pi r)^2}+ \frac{C}{r^{2}}
    \cos({2\kF} r)            \,.
\end{align}
Here, we have set $K_\rho=K_\sigma=1$, as appropriate for free fermions.  In
contrast, the Peierls state is formally described by $K_\rho=0$ (implying
long-range $2\kF$ charge order) and $K_\sigma=0$ (reflecting the spin gap
which leads to exponentially suppressed spin
correlations).\cite{PhysRevB.87.075149}

\begin{figure}
  \includegraphics[width=0.45\textwidth,clip]{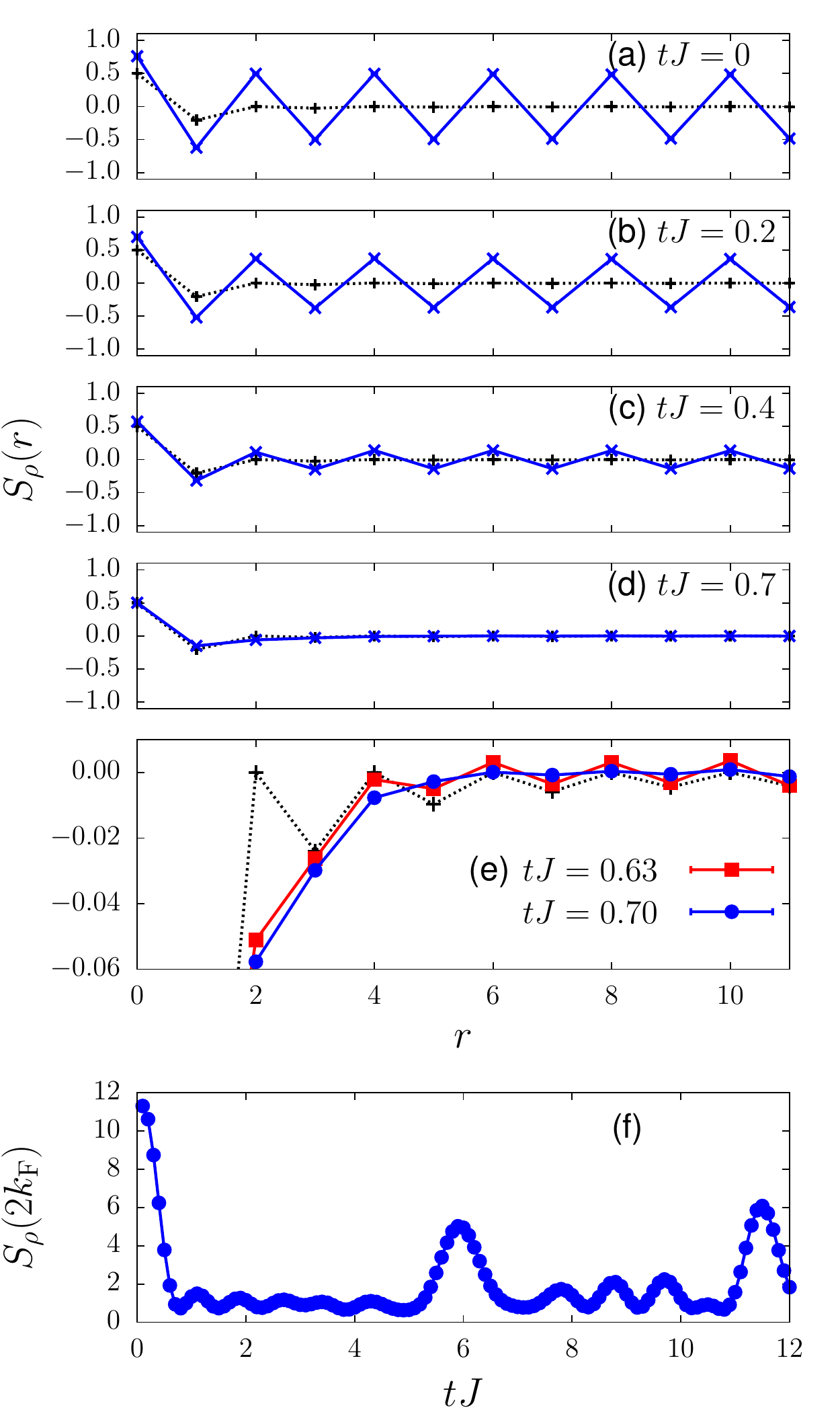}
  \caption{\label{fig:decaycharge}
    (Color online) Real-space charge correlations $S_\rho(r)$ (full lines) at different times $t$
    after the quench. Here, $\lambda=0.5$, $\om_0=0.5J$, $L=22$ and $\beta
    J=44$. Dotted lines correspond to $S_\rho(r)$ for $\lambda=0$. Panel (e)
    shows a closeup of panel (d), and additional results for $tJ=0.63$.
    (f) Time evolution of the amplitude for $q=2\kF$ charge correlations.
  }
\end{figure}

Figure~\ref{fig:decaycharge} shows results for $S_\rho(r)$ at different times
$t$ for $\om_0=0.5J$ and $\lambda=0.5$, as calculated from the time-dependent Green's
function~(\ref{eq:propG}). Panels (a)--(d) correspond to different times $tJ$
in the interval $[0,0.7]$, whereas panel (e) shows a closeup. Full lines
show the charge correlation function at time $t$ after the quench, whereas
dotted lines are results for free fermions obtained numerically.

We first consider the short-time behavior.
At $t=0$, Fig.~\ref{fig:decaycharge}(a), we observe the clear $2\kF$
charge correlations expected for the Peierls state. At $t>0$, after the quench
to $\lambda=0$, these oscillations decay over a time scale of about
$J/\sqrt{2}$. For $t$ close to this value, $2\kF$ oscillations are
practically absent at large distances. Shortly before that time, namely for
$tJ\approx0.63$, small-amplitude $2\kF$ oscillations comparable to the
free-fermion results are visible, see Fig.~\ref{fig:decaycharge}(e).  Whereas
the charge correlations for $tJ=0.7$ in Fig.~\ref{fig:decaycharge}(d) agree
well with those for the noninteracting system on the scale of the figure, the
closeup in Fig.~\ref{fig:decaycharge}(e) reveals that noticeable differences
remain at small distances even for the ``optimal time'' of $tJ=0.63$, which may
be expected in the absence of real thermalization in an integrable system. 

The results for the charge structure factor $S_\rho(q)$ (the Fourier
transform of the real-space charge correlator) at $q=2\kF$, shown in
Fig.~\ref{fig:decaycharge}(f), again illustrate the time scale  $tJ\approx0.7$
for the initial decay, as well as oscillations and partial revivals at larger
times. A detailed discussion of revivals will be given below for the case of
the double occupation, which shows the same overall features.

\begin{figure}
  \includegraphics[width=0.45\textwidth,clip]{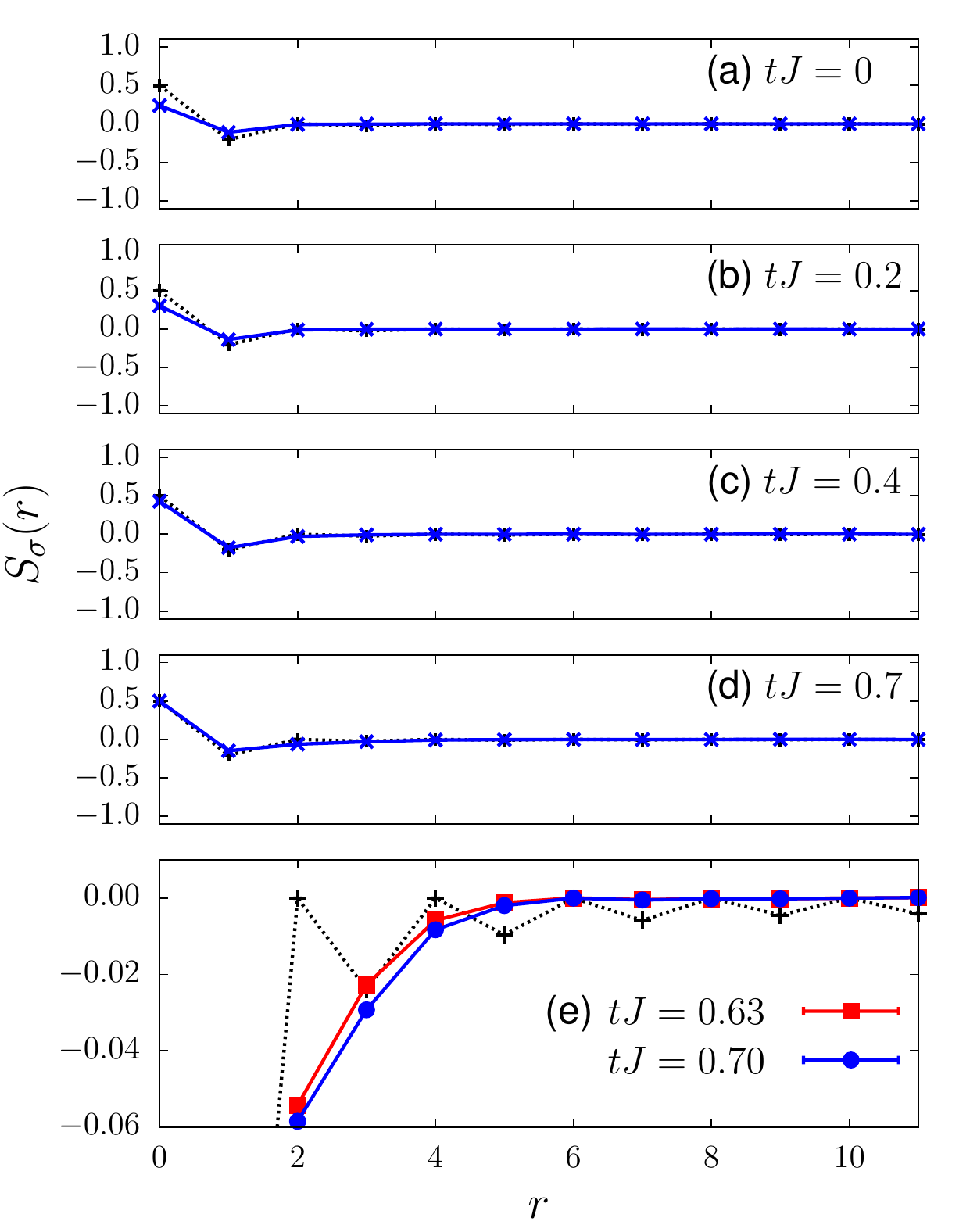}
  \caption{\label{fig:decayspin}
    (Color online) Real-space spin correlations $S_\sigma(r)$ (full lines) at different times $t$
    after the quench. Here, $\lambda=0.5$, $\om_0=0.5J$, $L=22$ and $\beta
    J=44$. Dotted lines correspond to $S_\sigma(r)$ for $\lambda=0$. Panel (e)
    shows a closeup of panel (d), and additional results for $tJ=0.63$.
  }
\end{figure}

Figure~\ref{fig:decayspin} shows the corresponding results for the spin
correlation function $S_\sigma(r)$.  The times shown are
the same as in Fig.~\ref{fig:decaycharge}. In contrast to $S_\rho(r)$,
Fig.~\ref{fig:decaycharge}(a), there are no discernible $2\kF$ correlations
at $t=0$, in accordance with the nonzero spin gap of the Peierls state. 
After the quench, the deviations from the noninteracting results visible at
short distances diminish as a function of time. For $tJ=0.7$, similar to
the charge correlations, $S_\sigma(r)$ looks very similar to the
free-fermion result on the scale of Fig.~\ref{fig:decayspin}(d),
but the closeup in Fig.~\ref{fig:decayspin}(e) again reveals clear
differences at small distances. Interestingly, the results for
$S_\sigma(r)$ in Fig.~\ref{fig:decayspin}(e) agree within the symbol
size used with those for $S_\rho(r)$ in Fig.~\ref{fig:decaycharge}(e),
despite substantial differences at $t=0$.

Because the system is quenched to a noninteracting and hence integrable
Hamiltonian $\hat{H}_0$, real thermalization is not
possible.\cite{PhysRevLett.98.050405} Instead, the time evolution seen in
Figs.~\ref{fig:decaycharge} and~\ref{fig:decayspin} is related to
dephasing. Conserved quantities such as the total energy and the momentum
distribution function are independent of $t$.  Because the time evolution is
determined by $\hat{H}_0$, it is independent of the electron-phonon
parameters $\lambda$ and $\omega_0$. However, the values of these parameters
do modify the initial state and hence have an impact on results at $t>0$. As
emphasized before, our method fully accounts for quantum phonon (and thermal)
fluctuations in the
initial state.  Most importantly, an increase of the phonon frequency brings
the initial state closer to the Peierls phase boundary.\cite{hardikar:245103} 

\begin{figure}[t]
  \includegraphics[width=0.45\textwidth,clip]{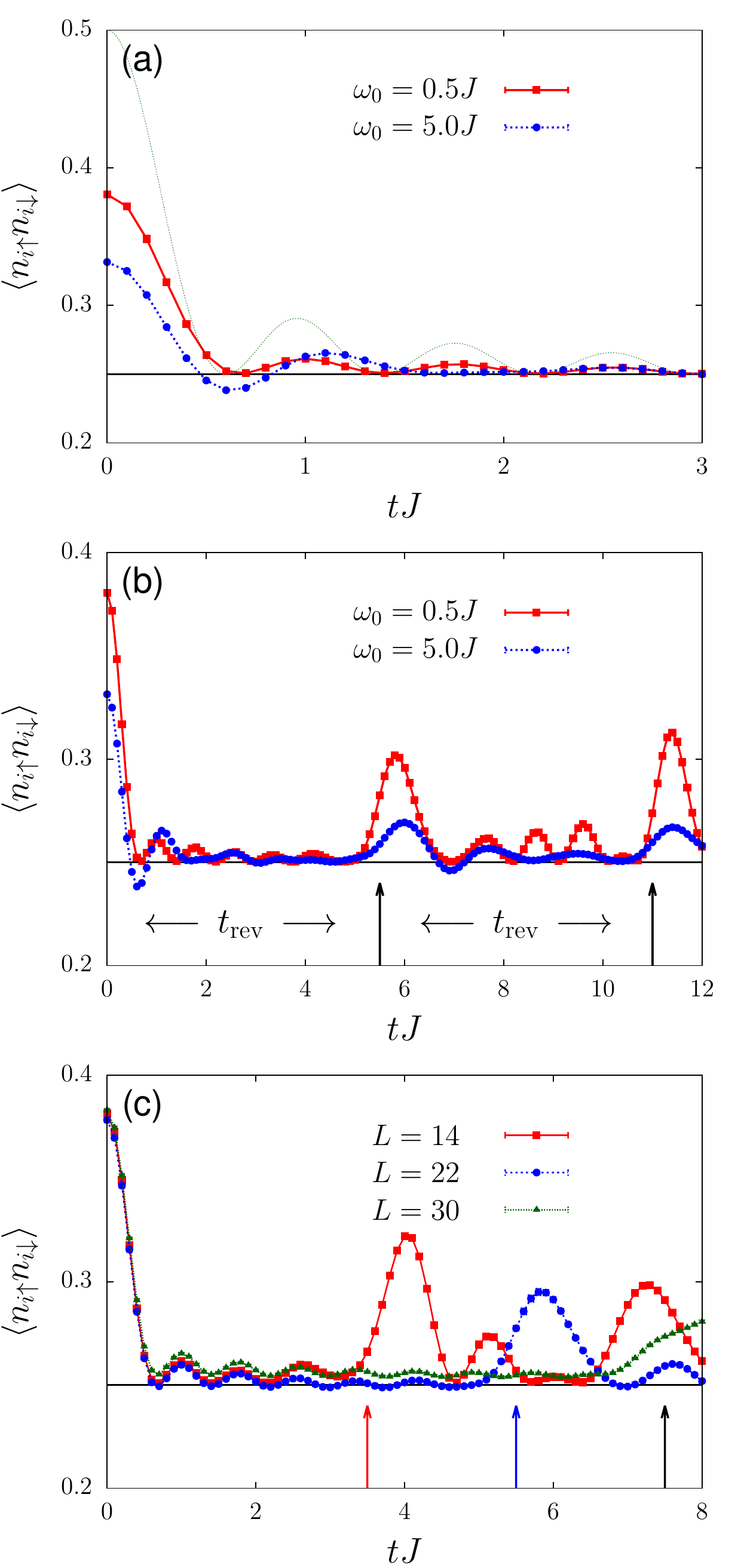}
  \caption{\label{fig:decaydocc}
    (Color online) Double occupation as a function of time at fixed
    $\lambda=0.5$.
    (a) Short-time behavior for two different phonon
    frequencies at $L=22$ and $\beta J=44$. The dotted line corresponds to
    the free-fermion result, see text, and the vertical line to the value
    0.25. 
    (b)
    As in (a) but showing a larger time interval. Arrows indicate the first
    two periods of the revival time $t_\text{rev} J=L/4$.
    (c) Results for different system sizes for $\omega_0=0.5J$ and $\beta
    J=L$. Arrows indicate $t_\text{rev}$.
  }
\end{figure}

To illustrate the dependence of charge order and its time evolution on the
phonon frequency, we show in Fig.~\ref{fig:decaydocc}(a) the double occupancy,
$\las n_{i\UP} n_{i\DO}\ras$, which reflects the pairing of electrons at the
same site in the Peierls state, and its decay after the quench. Comparing the
cases of $\om_0=0.5J$ and $\om_0=5J$, Fig.~\ref{fig:decaydocc}(a) reveals a
significantly smaller double occupation at $t=0$ for $\om_0=5J$ than for
$\om_0=0.5J$. At $t>0$, we observe a fast initial decay of the double
occupation toward the noninteracting value 0.25. The time scale is the same
as that for the decay of $2\kF$ charge correlations in
Fig.~\ref{fig:decaycharge}(f). After the first minimum, oscillations with a
frequency $\sim0.5/J$ occur, and the double occupation approaches 0.25.

To better understand these results, we also include in
Fig.~\ref{fig:decaydocc}(a) the result for the decay of the double occupation
in a free-fermion model, with an initial state corresponding to a perfect
alternation of empty and doubly occupied sites, $\ket{\Psi_\mathrm{CDW}}=\prod_i
c^\dag_{2i\UP}c^\dag_{2i\DO}\ket{0}$.\cite{EnnsSirker2012} This state may be
regarded as a perfectly dimerized, mean-field Peierls state. The time
evolution of the double occupation starting from $\ket{\Psi_\mathrm{CDW}}$ is given by\cite{EnnsSirker2012}
$\las \on_{i\UP} \on_{i\DO}\ras = \oq \left[1 + J_0^2(4tJ)\right]$,
where $J_0$ is the Bessel function of the first kind. Starting from the value
0.5 corresponding to full dimerization at $t=0$, the double occupation decays
toward 0.25, showing several periods of oscillation. This behavior is very
similar to the results for the Holstein model, and the time scale for the
initial decay agrees quite well, suggesting that the latter is related to the
noninteracting Hamiltonian. The double occupation always stays
above the noninteracting value 0.25 in the free-fermion results. Whereas the
same is true for the data for  $\om_0=0.5J$ in Fig.~\ref{fig:decaydocc}(a),
the results for $\om_0=5J$ fall below 0.25 at short times. A possible
explanation for this difference is that the case of $\om_0=0.5J$ is
closer to the mean-field limit $\om_0=0$. Finally, whereas the height of
consecutive maxima decreases monotonically for free fermions, this is not the
case for the Holstein model (the fourth
maximum is higher than the third for $\om_0=5J$). We attribute these
differences to correlations in the initial state of the Holstein model.

At larger times, shown in Fig.~\ref{fig:decaydocc}(b), we observe partial
revivals of the double occupation. The time between such revivals,
$t_\text{rev}$, is constant and independent of the initial state, as seen in
Fig.~\ref{fig:decaydocc}(b). This observation can be explained in terms of
the findings of Ref.~\onlinecite{PhysRevA.85.032114}; the revival time of an
integrable system after a quench is\cite{PhysRevA.85.032114}
\begin{equation}\label{eq:trev}
  t_\text{rev} \approx \frac{L}{2v_\text{max}}\,,
\end{equation}
where $v_\text{max}$ is the maximal velocity, in our case given by
$v_\text{max}=\max|\partial \epsilon(k)/\partial k|=2J$. Physically, $v_\text{max}$ is the maximum velocity for the
propagation of information after the quench. Using the result for
$v_\text{max}$, we expect $t_\text{rev}=L/4J = 5.5$, in very good agreement
with the revivals visible in Fig.~\ref{fig:decaydocc}(a). To further verify the
validity of Eq.~(\ref{eq:trev}), we show in Fig.~\ref{fig:decaydocc}(b) the
double occupation as a function of time for three different system sizes
$L$. Again, the estimate $t_\text{rev}=L/4J$ agrees well with the onset of
the first revival. The same revival times can also be identified in the
results for the charge correlations shown in Fig.~\ref{fig:decaycharge}(f).
Our observations are thus in agreement with the general statement in
Ref.~\onlinecite{PhysRevA.85.032114} that the revival time does not depend on
the initial state if the Hamiltonian is local, see also Ref.~\onlinecite{LiebRobinson72}.

\vspace*{1em}
\section{Conclusions}\label{sec:conclusions}

Motivated by an increasing number of experiments on ultrafast dynamics after
photo-induced phase transitions, and by the enormous challenge of numerically
describing time-dependent phenomena in systems with electron-phonon
interaction, we considered the quench of a Peierls insulator to a
noninteracting Hamiltonian. Using an extension of the continuous-time quantum
Monte Carlo method, we presented exact results for equal-time charge and spin
correlation functions. The time evolution of the latter is related to
dephasing, and we find that the dominant $2\kF$ charge correlations of the
initial Peierls state decay over a time scale of about $J/\sqrt{2}$. In the
absence of thermalization, differences to the correlation functions of free
fermions remain. At longer times, we observe partial revivals, with revival
times that depend on the maximal velocity and system size.

In the present case, the electron-phonon coupling and the quantum nature of
phonons only play a role for the initial state. Increasing (decreasing) the
phonon frequency (the coupling) moves the system closer to the metallic
state, thereby suppressing charge order.  Importantly, the results shown
cannot be obtained by considering the antiadiabatic limit $\om_0\to\infty$,
where the Holstein model maps onto an attractive Hubbard model. The reason is
that the attractive Hubbard model at half filling always remains metallic,
with the Luttinger liquid parameter $K_\rho$ fixed to one by symmetry.\cite{Giamarchi}

The exact results presented here can serve as benchmarks for more realistic
calculations using, for example, the density-matrix renormalization group.
In the latter, the phonon Hilbert space has to be truncated for the initial
state as well as for the time evolution. Finally, our findings may even
become directly relevant for experiments with cold atoms or trapped ions;
several different proposals exist how to realize electron-phonon Hamiltonians
in such systems.\cite{PhysRevA.84.051401,1367-2630-14-3-033019,PhysRevA.72.033609,PhysRevLett.109.200501,PhysRevLett.109.250501}

{\begin{acknowledgments}%
Computer time at the J\"ulich Supercomputing Centre, support from
the DFG Grant No.~Ho~4489/2-1 (Forschergruppe FOR 1807), and discussions with
F. F. Assaad, F. Goth, and M. Sentef are gratefully acknowledged.
\end{acknowledgments}}

%\bibliography{../refs}

\begin{thebibliography}{10}%
\makeatletter
\providecommand \@ifxundefined [1]{%
 \ifx #1\undefined \expandafter \@firstoftwo
 \else \expandafter \@secondoftwo
\fi
}%
\providecommand \@ifnum [1]{%
 \ifnum #1\expandafter \@firstoftwo
 \else \expandafter \@secondoftwo
\fi
}%
\providecommand \enquote [1]{``#1''}%
\providecommand \bibnamefont  [1]{#1}%
\providecommand \bibfnamefont [1]{#1}%
\providecommand \citenamefont [1]{#1}%
\providecommand\href[0]{\@sanitize\@href}%
\providecommand\@href[1]{\endgroup\@@startlink{#1}\endgroup\@@href}%
\providecommand\@@href[1]{#1\@@endlink}%
\providecommand \@sanitize [0]{\begingroup\catcode`\&12\catcode`\#12\relax}%
\@ifxundefined \pdfoutput {\@firstoftwo}{%
 \@ifnum{\z@=\pdfoutput}{\@firstoftwo}{\@secondoftwo}%
}{%
 \providecommand\@@startlink[1]{\leavevmode}%
 \providecommand\@@endlink[0]{}%
}{%
 \providecommand\@@startlink[1]{%
  \leavevmode
  \pdfstartlink
   attr{/Border[0 0 1 ]/H/I/C[0 1 1]}%
   user{/Subtype/Link/A<</Type/Action/S/URI/URI(#1)>>}%
  \relax
 }%
 \providecommand\@@endlink[0]{\pdfendlink}%
}%
\providecommand \url  [0]{\begingroup\@sanitize \@url }%
\providecommand \@url [1]{\endgroup\@href {#1}{\urlprefix}}%
\providecommand \urlprefix [0]{URL }%
\providecommand \Eprint[0]{\href }%
\@ifxundefined \urlstyle {%
  \providecommand \doi [1]{doi:\discretionary{}{}{}#1}%
}{%
  \providecommand \doi [0]{doi:\discretionary{}{}{}\begingroup
  \urlstyle{rm}\Url }%
}%
\providecommand \doibase [0]{http://dx.doi.org/}%
\providecommand \Doi[1]{\href{\doibase#1}}%
\providecommand \bibAnnote [3]{%
  \BibitemShut{#1}%
  \begin{quotation}\noindent
    \textsc{Key:}\ #2\\\textsc{Annotation:}\ #3%
  \end{quotation}%
}%
\providecommand \bibAnnoteFile [2]{%
  \IfFileExists{#2}{\bibAnnote {#1} {#2} {\input{#2}}}{}%
}%
\providecommand \typeout [0]{\immediate \write \m@ne }%
\providecommand \selectlanguage [0]{\@gobble}%
\providecommand \bibinfo [0]{\@secondoftwo}%
\providecommand \bibfield [0]{\@secondoftwo}%
\providecommand \translation [1]{[#1]}%
\providecommand \BibitemOpen[0]{}%
\providecommand \bibitemStop [0]{}%
\providecommand \bibitemNoStop [0]{.\EOS\space}%
\providecommand \EOS [0]{\spacefactor3000\relax}%
\providecommand \BibitemShut [1]{\csname bibitem#1\endcsname}%
%</preamble>
\bibitem{Greiner00}%
  \BibitemOpen
  \bibfield{author}{%
  \bibinfo {author} {\bibfnamefont{M.}~\bibnamefont{Greiner}}, \bibinfo
  {author} {\bibfnamefont{O.}~\bibnamefont{Mandel}}, \bibinfo {author}
  {\bibfnamefont{T.}~\bibnamefont{Esslinger}}, \bibinfo {author}
  {\bibfnamefont{T.~W.}\ \bibnamefont{H\"ansch}},\ and\ \bibinfo {author}
  {\bibfnamefont{I.}~\bibnamefont{Bloch}},\ }%
  \bibfield{journal}{%
  \bibinfo {journal} {Nature (London)}\ }%
  \textbf{\bibinfo {volume} {415}},\ \bibinfo {pages} {39} (\bibinfo {year}
  {2002})%
  \bibAnnoteFile{NoStop}{Greiner00}%
\bibitem{PhysRevLett.105.067001}%
  \BibitemOpen
  \bibfield{author}{%
  \bibinfo {author} {\bibfnamefont{A.}~\bibnamefont{Pashkin}}, \bibinfo
  {author} {\bibfnamefont{M.}~\bibnamefont{Porer}}, \bibinfo {author}
  {\bibfnamefont{M.}~\bibnamefont{Beyer}}, \bibinfo {author}
  {\bibfnamefont{K.~W.}\ \bibnamefont{Kim}}, \bibinfo {author}
  {\bibfnamefont{A.}~\bibnamefont{Dubroka}}, \bibinfo {author}
  {\bibfnamefont{C.}~\bibnamefont{Bernhard}}, \bibinfo {author}
  {\bibfnamefont{X.}~\bibnamefont{Yao}}, \bibinfo {author}
  {\bibfnamefont{Y.}~\bibnamefont{Dagan}}, \bibinfo {author}
  {\bibfnamefont{R.}~\bibnamefont{Hackl}}, \bibinfo {author}
  {\bibfnamefont{A.}~\bibnamefont{Erb}}, \bibinfo {author}
  {\bibfnamefont{J.}~\bibnamefont{Demsar}}, \bibinfo {author}
  {\bibfnamefont{R.}~\bibnamefont{Huber}},\ and\ \bibinfo {author}
  {\bibfnamefont{A.}~\bibnamefont{Leitenstorfer}},\ }%
  \bibfield{journal}{%
  \Doi{10.1103/PhysRevLett.105.067001}{\bibinfo {journal} {Phys. Rev. Lett.}}\
  }%
  \textbf{\bibinfo {volume} {105}},\ \bibinfo {pages} {067001} (\bibinfo {year}
  {2010})%
  \bibAnnoteFile{NoStop}{PhysRevLett.105.067001}%
\bibitem{PhysRevLett.105.257001}%
  \BibitemOpen
  \bibfield{author}{%
  \bibinfo {author} {\bibfnamefont{C.}~\bibnamefont{Gadermaier}}, \bibinfo
  {author} {\bibfnamefont{A.~S.}\ \bibnamefont{Alexandrov}}, \bibinfo {author}
  {\bibfnamefont{V.~V.}\ \bibnamefont{Kabanov}}, \bibinfo {author}
  {\bibfnamefont{P.}~\bibnamefont{Kusar}}, \bibinfo {author}
  {\bibfnamefont{T.}~\bibnamefont{Mertelj}}, \bibinfo {author}
  {\bibfnamefont{X.}~\bibnamefont{Yao}}, \bibinfo {author}
  {\bibfnamefont{C.}~\bibnamefont{Manzoni}}, \bibinfo {author}
  {\bibfnamefont{D.}~\bibnamefont{Brida}}, \bibinfo {author}
  {\bibfnamefont{G.}~\bibnamefont{Cerullo}},\ and\ \bibinfo {author}
  {\bibfnamefont{D.}~\bibnamefont{Mihailovic}},\ }%
  \bibfield{journal}{%
  \Doi{10.1103/PhysRevLett.105.257001}{\bibinfo {journal} {Phys. Rev. Lett.}}\
  }%
  \textbf{\bibinfo {volume} {105}},\ \bibinfo {pages} {257001} (\bibinfo {year}
  {2010})%
  \bibAnnoteFile{NoStop}{PhysRevLett.105.257001}%
\bibitem{PhysRevLett.105.246402}%
  \BibitemOpen
  \bibfield{author}{%
  \bibinfo {author} {\bibfnamefont{Y.}~\bibnamefont{Kawakami}}, \bibinfo
  {author} {\bibfnamefont{T.}~\bibnamefont{Fukatsu}}, \bibinfo {author}
  {\bibfnamefont{Y.}~\bibnamefont{Sakurai}}, \bibinfo {author}
  {\bibfnamefont{H.}~\bibnamefont{Unno}}, \bibinfo {author}
  {\bibfnamefont{H.}~\bibnamefont{Itoh}}, \bibinfo {author}
  {\bibfnamefont{S.}~\bibnamefont{Iwai}}, \bibinfo {author}
  {\bibfnamefont{T.}~\bibnamefont{Sasaki}}, \bibinfo {author}
  {\bibfnamefont{K.}~\bibnamefont{Yamamoto}}, \bibinfo {author}
  {\bibfnamefont{K.}~\bibnamefont{Yakushi}},\ and\ \bibinfo {author}
  {\bibfnamefont{K.}~\bibnamefont{Yonemitsu}},\ }%
  \bibfield{journal}{%
  \Doi{10.1103/PhysRevLett.105.246402}{\bibinfo {journal} {Phys. Rev. Lett.}}\
  }%
  \textbf{\bibinfo {volume} {105}},\ \bibinfo {pages} {246402} (\bibinfo {year}
  {2010})%
  \bibAnnoteFile{NoStop}{PhysRevLett.105.246402}%
\bibitem{Wall2011}%
  \BibitemOpen
  \bibfield{author}{%
  \bibinfo {author} {\bibfnamefont{S.}~\bibnamefont{Wall}}, \bibinfo {author}
  {\bibfnamefont{D.}~\bibnamefont{Brida}}, \bibinfo {author}
  {\bibfnamefont{S.~R.}\ \bibnamefont{Clark}}, \bibinfo {author}
  {\bibfnamefont{H.~P.}\ \bibnamefont{Ehrke}}, \bibinfo {author}
  {\bibfnamefont{D.}~\bibnamefont{Jaksch}}, \bibinfo {author}
  {\bibfnamefont{A.}~\bibnamefont{Ardavan}}, \bibinfo {author}
  {\bibfnamefont{S.}~\bibnamefont{Bonora}}, \bibinfo {author}
  {\bibfnamefont{H.}~\bibnamefont{Uemura}}, \bibinfo {author}
  {\bibfnamefont{Y.}~\bibnamefont{Takahashi}}, \bibinfo {author}
  {\bibfnamefont{T.}~\bibnamefont{Hasegawa}}, \bibinfo {author}
  {\bibfnamefont{H.}~\bibnamefont{Okamoto}}, \bibinfo {author}
  {\bibfnamefont{G.}~\bibnamefont{Cerullo}},\ and\ \bibinfo {author}
  {\bibfnamefont{A.}~\bibnamefont{Cavalleri}},\ }%
  \bibfield{journal}{%
  \bibinfo {journal} {Nature Phys.}\ }%
  \textbf{\bibinfo {volume} {7}},\ \bibinfo {pages} {114} (\bibinfo {year}
  {2011})%
  \bibAnnoteFile{NoStop}{Wall2011}%
\bibitem{PhysRevLett.99.197001}%
  \BibitemOpen
  \bibfield{author}{%
  \bibinfo {author} {\bibfnamefont{L.}~\bibnamefont{Perfetti}}, \bibinfo
  {author} {\bibfnamefont{P.~A.}\ \bibnamefont{Loukakos}}, \bibinfo {author}
  {\bibfnamefont{M.}~\bibnamefont{Lisowski}}, \bibinfo {author}
  {\bibfnamefont{U.}~\bibnamefont{Bovensiepen}}, \bibinfo {author}
  {\bibfnamefont{H.}~\bibnamefont{Eisaki}},\ and\ \bibinfo {author}
  {\bibfnamefont{M.}~\bibnamefont{Wolf}},\ }%
  \bibfield{journal}{%
  \Doi{10.1103/PhysRevLett.99.197001}{\bibinfo {journal} {Phys. Rev. Lett.}}\
  }%
  \textbf{\bibinfo {volume} {99}},\ \bibinfo {pages} {197001} (\bibinfo {year}
  {2007})%
  \bibAnnoteFile{NoStop}{PhysRevLett.99.197001}%
\bibitem{PhysRevLett.107.097002}%
  \BibitemOpen
  \bibfield{author}{%
  \bibinfo {author} {\bibfnamefont{R.}~\bibnamefont{Cort\'es}}, \bibinfo
  {author} {\bibfnamefont{L.}~\bibnamefont{Rettig}}, \bibinfo {author}
  {\bibfnamefont{Y.}~\bibnamefont{Yoshida}}, \bibinfo {author}
  {\bibfnamefont{H.}~\bibnamefont{Eisaki}}, \bibinfo {author}
  {\bibfnamefont{M.}~\bibnamefont{Wolf}},\ and\ \bibinfo {author}
  {\bibfnamefont{U.}~\bibnamefont{Bovensiepen}},\ }%
  \bibfield{journal}{%
  \Doi{10.1103/PhysRevLett.107.097002}{\bibinfo {journal} {Phys. Rev. Lett.}}\
  }%
  \textbf{\bibinfo {volume} {107}},\ \bibinfo {pages} {097002} (\bibinfo {year}
  {2011})%
  \bibAnnoteFile{NoStop}{PhysRevLett.107.097002}%
\bibitem{Chollet07012005}%
  \BibitemOpen
  \bibfield{author}{%
  \bibinfo {author} {\bibfnamefont{M.}~\bibnamefont{Chollet}}, \bibinfo
  {author} {\bibfnamefont{L.}~\bibnamefont{Guerin}}, \bibinfo {author}
  {\bibfnamefont{N.}~\bibnamefont{Uchida}}, \bibinfo {author}
  {\bibfnamefont{S.}~\bibnamefont{Fukaya}}, \bibinfo {author}
  {\bibfnamefont{H.}~\bibnamefont{Shimoda}}, \bibinfo {author}
  {\bibfnamefont{T.}~\bibnamefont{Ishikawa}}, \bibinfo {author}
  {\bibfnamefont{K.}~\bibnamefont{Matsuda}}, \bibinfo {author}
  {\bibfnamefont{T.}~\bibnamefont{Hasegawa}}, \bibinfo {author}
  {\bibfnamefont{A.}~\bibnamefont{Ota}}, \bibinfo {author}
  {\bibfnamefont{H.}~\bibnamefont{Yamochi}}, \bibinfo {author}
  {\bibfnamefont{G.}~\bibnamefont{Saito}}, \bibinfo {author}
  {\bibfnamefont{R.}~\bibnamefont{Tazaki}}, \bibinfo {author}
  {\bibfnamefont{S.-i.}\ \bibnamefont{Adachi}},\ and\ \bibinfo {author}
  {\bibfnamefont{S.-y.}\ \bibnamefont{Koshihara}},\ }%
  \bibfield{journal}{%
  \Doi{10.1126/science.1105067}{\bibinfo {journal} {Science}}\ }%
  \textbf{\bibinfo {volume} {307}},\ \bibinfo {pages} {86} (\bibinfo {year}
  {2005})%
  \bibAnnoteFile{NoStop}{Chollet07012005}%
\bibitem{PhysRevLett.98.097402}%
  \BibitemOpen
  \bibfield{author}{%
  \bibinfo {author} {\bibfnamefont{S.}~\bibnamefont{Iwai}}, \bibinfo {author}
  {\bibfnamefont{K.}~\bibnamefont{Yamamoto}}, \bibinfo {author}
  {\bibfnamefont{A.}~\bibnamefont{Kashiwazaki}}, \bibinfo {author}
  {\bibfnamefont{F.}~\bibnamefont{Hiramatsu}}, \bibinfo {author}
  {\bibfnamefont{H.}~\bibnamefont{Nakaya}}, \bibinfo {author}
  {\bibfnamefont{Y.}~\bibnamefont{Kawakami}}, \bibinfo {author}
  {\bibfnamefont{K.}~\bibnamefont{Yakushi}}, \bibinfo {author}
  {\bibfnamefont{H.}~\bibnamefont{Okamoto}}, \bibinfo {author}
  {\bibfnamefont{H.}~\bibnamefont{Mori}},\ and\ \bibinfo {author}
  {\bibfnamefont{Y.}~\bibnamefont{Nishio}},\ }%
  \bibfield{journal}{%
  \Doi{10.1103/PhysRevLett.98.097402}{\bibinfo {journal} {Phys. Rev. Lett.}}\
  }%
  \textbf{\bibinfo {volume} {98}},\ \bibinfo {pages} {097402} (\bibinfo {year}
  {2007})%
  \bibAnnoteFile{NoStop}{PhysRevLett.98.097402}%
\bibitem{PhysRevLett.99.116401}%
  \BibitemOpen
  \bibfield{author}{%
  \bibinfo {author} {\bibfnamefont{C.}~\bibnamefont{K\"ubler}}, \bibinfo
  {author} {\bibfnamefont{H.}~\bibnamefont{Ehrke}}, \bibinfo {author}
  {\bibfnamefont{R.}~\bibnamefont{Huber}}, \bibinfo {author}
  {\bibfnamefont{R.}~\bibnamefont{Lopez}}, \bibinfo {author}
  {\bibfnamefont{A.}~\bibnamefont{Halabica}}, \bibinfo {author}
  {\bibfnamefont{R.~F.}\ \bibnamefont{Haglund}},\ and\ \bibinfo {author}
  {\bibfnamefont{A.}~\bibnamefont{Leitenstorfer}},\ }%
  \bibfield{journal}{%
  \Doi{10.1103/PhysRevLett.99.116401}{\bibinfo {journal} {Phys. Rev. Lett.}}\
  }%
  \textbf{\bibinfo {volume} {99}},\ \bibinfo {pages} {116401} (\bibinfo {year}
  {2007})%
  \bibAnnoteFile{NoStop}{PhysRevLett.99.116401}%
\bibitem{PhysRevLett.102.066404}%
  \BibitemOpen
  \bibfield{author}{%
  \bibinfo {author} {\bibfnamefont{A.}~\bibnamefont{Tomeljak}}, \bibinfo
  {author} {\bibfnamefont{H.}~\bibnamefont{Sch\"afer}}, \bibinfo {author}
  {\bibfnamefont{D.}~\bibnamefont{St\"adter}}, \bibinfo {author}
  {\bibfnamefont{M.}~\bibnamefont{Beyer}}, \bibinfo {author}
  {\bibfnamefont{K.}~\bibnamefont{Biljakovic}},\ and\ \bibinfo {author}
  {\bibfnamefont{J.}~\bibnamefont{Demsar}},\ }%
  \bibfield{journal}{%
  \Doi{10.1103/PhysRevLett.102.066404}{\bibinfo {journal} {Phys. Rev. Lett.}}\
  }%
  \textbf{\bibinfo {volume} {102}},\ \bibinfo {pages} {066404} (\bibinfo {year}
  {2009})%
  \bibAnnoteFile{NoStop}{PhysRevLett.102.066404}%
\bibitem{Hellmann2010}%
  \BibitemOpen
  \bibfield{author}{%
  \bibinfo {author} {\bibfnamefont{S.}~\bibnamefont{Hellmann}}, \bibinfo
  {author} {\bibfnamefont{M.}~\bibnamefont{Beye}}, \bibinfo {author}
  {\bibfnamefont{C.}~\bibnamefont{Sohrt}}, \bibinfo {author}
  {\bibfnamefont{T.}~\bibnamefont{Rohwer}}, \bibinfo {author}
  {\bibfnamefont{F.}~\bibnamefont{Sorgenfrei}}, \bibinfo {author}
  {\bibfnamefont{H.}~\bibnamefont{Redlin}}, \bibinfo {author}
  {\bibfnamefont{M.}~\bibnamefont{Kall\"ane}}, \bibinfo {author}
  {\bibfnamefont{M.}~\bibnamefont{Marczynski-B\"uhlow}}, \bibinfo {author}
  {\bibfnamefont{F.}~\bibnamefont{Hennies}}, \bibinfo {author}
  {\bibfnamefont{M.}~\bibnamefont{Bauer}}, \bibinfo {author}
  {\bibfnamefont{A.}~\bibnamefont{F\"ohlisch}}, \bibinfo {author}
  {\bibfnamefont{L.}~\bibnamefont{Kipp}}, \bibinfo {author}
  {\bibfnamefont{W.}~\bibnamefont{Wurth}},\ and\ \bibinfo {author}
  {\bibfnamefont{K.}~\bibnamefont{Rossnagel}},\ }%
  \bibfield{journal}{%
  \Doi{10.1103/PhysRevLett.105.187401}{\bibinfo {journal} {Phys. Rev. Lett.}}\ }%
  \textbf{\bibinfo {volume} {105}},\ \bibinfo {pages} {187401} (\bibinfo {year}
  {2010})%
  \bibAnnoteFile{NoStop}{Hellmann2010}%
\bibitem{PhysRevB.81.073102}%
  \BibitemOpen
  \bibfield{author}{%
  \bibinfo {author} {\bibfnamefont{R.~G.}\ \bibnamefont{Moore}}, \bibinfo
  {author} {\bibfnamefont{V.}~\bibnamefont{Brouet}}, \bibinfo {author}
  {\bibfnamefont{R.}~\bibnamefont{He}}, \bibinfo {author}
  {\bibfnamefont{D.~H.}\ \bibnamefont{Lu}}, \bibinfo {author}
  {\bibfnamefont{N.}~\bibnamefont{Ru}}, \bibinfo {author}
  {\bibfnamefont{J.-H.}\ \bibnamefont{Chu}}, \bibinfo {author}
  {\bibfnamefont{I.~R.}\ \bibnamefont{Fisher}},\ and\ \bibinfo {author}
  {\bibfnamefont{Z.-X.}\ \bibnamefont{Shen}},\ }%
  \bibfield{journal}{%
  \Doi{10.1103/PhysRevB.81.073102}{\bibinfo {journal} {Phys. Rev. B}}\ }%
  \textbf{\bibinfo {volume} {81}},\ \bibinfo {pages} {073102} (\bibinfo {year}
  {2010})%
  \bibAnnoteFile{NoStop}{PhysRevB.81.073102}%
\bibitem{JPSJ.79.034708}%
  \BibitemOpen
  \bibfield{author}{%
  \bibinfo {author} {\bibfnamefont{S.}~\bibnamefont{Miyashita}}, \bibinfo
  {author} {\bibfnamefont{Y.}~\bibnamefont{Tanaka}}, \bibinfo {author}
  {\bibfnamefont{S.}~\bibnamefont{Iwai}},\ and\ \bibinfo {author}
  {\bibfnamefont{K.}~\bibnamefont{Yonemitsu}},\ }%
  \bibfield{journal}{%
  \Doi{10.1143/JPSJ.79.034708}{\bibinfo {journal} {J. Phys. Soc. Jpn.}}\ }%
  \textbf{\bibinfo {volume} {79}},\ \bibinfo {pages} {034708} (\bibinfo {year}
  {2010})%
  \bibAnnoteFile{NoStop}{JPSJ.79.034708}%
\bibitem{Schollwock201196}%
  \BibitemOpen
  \bibfield{author}{%
  \bibinfo {author} {\bibfnamefont{U.}~\bibnamefont{Schollw\"ock}},\ }%
  \bibfield{journal}{%
  \Doi{10.1016/j.aop.2010.09.012}{\bibinfo {journal} {Annals of Physics}}\ }%
  \textbf{\bibinfo {volume} {326}},\ \bibinfo {pages} {96} (\bibinfo {year}
  {2011})%
  \bibAnnoteFile{NoStop}{Schollwock201196}%
\bibitem{SchmidtMonien02}%
  \BibitemOpen
  \bibfield{author}{%
  \bibinfo {author} {\bibfnamefont{P.}~\bibnamefont{Schmidt}}\ and\ \bibinfo
  {author} {\bibfnamefont{H.}~\bibnamefont{Monien}},\ }%
  \bibfield{journal}{%
  \bibinfo {journal} {cond-mat/0202046}}%
  \bibAnnoteFile{NoStop}{SchmidtMonien02}%
\bibitem{PhysRevLett.97.266408}%
  \BibitemOpen
  \bibfield{author}{%
  \bibinfo {author} {\bibfnamefont{J.~K.}\ \bibnamefont{Freericks}}, \bibinfo
  {author} {\bibfnamefont{V.~M.}\ \bibnamefont{Turkowski}},\ and\ \bibinfo
  {author} {\bibfnamefont{V.}~\bibnamefont{Zlati\ifmmode~\acute{c}\else
  \'{c}\fi{}}},\ }%
  \bibfield{journal}{%
  \Doi{10.1103/PhysRevLett.97.266408}{\bibinfo {journal} {Phys. Rev. Lett.}}\
  }%
  \textbf{\bibinfo {volume} {97}},\ \bibinfo {pages} {266408} (\bibinfo {year}
  {2006})%
  \bibAnnoteFile{NoStop}{PhysRevLett.97.266408}%
\bibitem{PhysRevB.84.115145}%
  \BibitemOpen
  \bibfield{author}{%
  \bibinfo {author} {\bibfnamefont{M.}~\bibnamefont{Knap}}, \bibinfo {author}
  {\bibfnamefont{W.}~\bibnamefont{von~der Linden}},\ and\ \bibinfo {author}
  {\bibfnamefont{E.}~\bibnamefont{Arrigoni}},\ }%
  \bibfield{journal}{%
  \Doi{10.1103/PhysRevB.84.115145}{\bibinfo {journal} {Phys. Rev. B}}\ }%
  \textbf{\bibinfo {volume} {84}},\ \bibinfo {pages} {115145} (\bibinfo {year}
  {2011})%
  \bibAnnoteFile{NoStop}{PhysRevB.84.115145}%
\bibitem{PhysRevLett.109.126406}%
  \BibitemOpen
  \bibfield{author}{%
  \bibinfo {author} {\bibfnamefont{C.}~\bibnamefont{Karrasch}}, \bibinfo
  {author} {\bibfnamefont{J.}~\bibnamefont{Rentrop}}, \bibinfo {author}
  {\bibfnamefont{D.}~\bibnamefont{Schuricht}},\ and\ \bibinfo {author}
  {\bibfnamefont{V.}~\bibnamefont{Meden}},\ }%
  \bibfield{journal}{%
  \Doi{10.1103/PhysRevLett.109.126406}{\bibinfo {journal} {Phys. Rev. Lett.}}\
  }%
  \textbf{\bibinfo {volume} {109}},\ \bibinfo {pages} {126406} (\bibinfo {year}
  {2012})%
  \bibAnnoteFile{NoStop}{PhysRevLett.109.126406}%
\bibitem{Ecksteinreview}%
  \BibitemOpen
  \bibfield{author}{%
  \bibinfo {author} {\bibfnamefont{M.}~\bibnamefont{Eckstein}}, \bibinfo
  {author} {\bibfnamefont{A.}~\bibnamefont{Hackl}}, \bibinfo {author}
  {\bibfnamefont{S.}~\bibnamefont{Kehrein}}, \bibinfo {author}
  {\bibfnamefont{M.}~\bibnamefont{Kollar}}, \bibinfo {author}
  {\bibfnamefont{M.}~\bibnamefont{Moeckel}}, \bibinfo {author}
  {\bibfnamefont{P.}~\bibnamefont{Werner}},\ and\ \bibinfo {author}
  {\bibfnamefont{F.}~\bibnamefont{Wolf}},\ }%
  \bibfield{journal}{%
  \Doi{10.1140/epjst/e2010-01219-x}{\bibinfo {journal} {Eur. Phys. J. Special Topics}}\ }%
  \textbf{\bibinfo {volume} {180}},\ \bibinfo {pages} {217} (\bibinfo {year}
  {2009})%
  \bibAnnoteFile{NoStop}{Ecksteinreview}%
\bibitem{Becca2013}%
  \BibitemOpen
  \bibfield{author}{%
  \bibinfo {author} {\bibfnamefont{E.}~\bibnamefont{Coira}}, \bibinfo {author}
  {\bibfnamefont{F.}~\bibnamefont{Becca}},\ and\ \bibinfo {author}
  {\bibfnamefont{A.}~\bibnamefont{Parola}},\ }%
  \bibfield{journal}{%
  \bibinfo {journal} {Eur. Phys. J. B}\ }%
  \textbf{\bibinfo {volume} {86}},\ \bibinfo {pages} {55} (\bibinfo {year}
  {2013})%
  \bibAnnoteFile{NoStop}{Becca2013}%
\bibitem{arXiv:1212.4841}%
  \BibitemOpen
  \bibfield{author}{%
  \bibinfo {author} {\bibfnamefont{M.}~\bibnamefont{Sentef}}, \bibinfo {author}
  {\bibfnamefont{A.~F.}\ \bibnamefont{Kemper}}, \bibinfo {author}
  {\bibfnamefont{B.}~\bibnamefont{Moritz}}, \bibinfo {author}
  {\bibfnamefont{J.~K.}\ \bibnamefont{Freericks}}, \bibinfo {author}
  {\bibfnamefont{Z.-X.}\ \bibnamefont{Shen}},\ and\ \bibinfo {author}
  {\bibfnamefont{T.~P.}\ \bibnamefont{Devereaux}},\ }%
  \bibfield{journal}{%
  \bibinfo {journal} {arXiv:1212.4841}}%
   (\bibinfo {year} {2012})%
  \bibAnnoteFile{NoStop}{arXiv:1212.4841}%
\bibitem{PhysRevLett.109.236402}%
  \BibitemOpen
  \bibfield{author}{%
  \bibinfo {author} {\bibfnamefont{D.}~\bibnamefont{Gole\ifmmode~\check{z}\else
  \v{z}\fi{}}}, \bibinfo {author}
  {\bibfnamefont{J.}~\bibnamefont{Bon\ifmmode~\check{c}\else \v{c}\fi{}a}},
  \bibinfo {author} {\bibfnamefont{L.}~\bibnamefont{Vidmar}},\ and\ \bibinfo
  {author} {\bibfnamefont{S.~A.}\ \bibnamefont{Trugman}},\ }%
  \bibfield{journal}{%
  \Doi{10.1103/PhysRevLett.109.236402}{\bibinfo {journal} {Phys. Rev. Lett.}}\
  }%
  \textbf{\bibinfo {volume} {109}},\ \bibinfo {pages} {236402} (\bibinfo {year}
  {2012})%
  \bibAnnoteFile{NoStop}{PhysRevLett.109.236402}%
\bibitem{PhysRevB.75.014307}%
  \BibitemOpen
  \bibfield{author}{%
  \bibinfo {author} {\bibfnamefont{L.-C.}\ \bibnamefont{Ku}}\ and\ \bibinfo
  {author} {\bibfnamefont{S.~A.}\ \bibnamefont{Trugman}},\ }%
  \bibfield{journal}{%
  \Doi{10.1103/PhysRevB.75.014307}{\bibinfo {journal} {Phys. Rev. B}}\ }%
  \textbf{\bibinfo {volume} {75}},\ \bibinfo {pages} {014307} (\bibinfo {year}
  {2007})%
  \bibAnnoteFile{NoStop}{PhysRevB.75.014307}%
\bibitem{PhysRevB.83.075104}%
  \BibitemOpen
  \bibfield{author}{%
  \bibinfo {author} {\bibfnamefont{H.}~\bibnamefont{Fehske}}, \bibinfo {author}
  {\bibfnamefont{G.}~\bibnamefont{Wellein}},\ and\ \bibinfo {author}
  {\bibfnamefont{A.~R.}\ \bibnamefont{Bishop}},\ }%
  \bibfield{journal}{%
  \Doi{10.1103/PhysRevB.83.075104}{\bibinfo {journal} {Phys. Rev. B}}\ }%
  \textbf{\bibinfo {volume} {83}},\ \bibinfo {pages} {075104} (\bibinfo {year}
  {2011})%
  \bibAnnoteFile{NoStop}{PhysRevB.83.075104}%
\bibitem{PhysRevB.85.144304}%
  \BibitemOpen
  \bibfield{author}{%
  \bibinfo {author} {\bibfnamefont{D.}~\bibnamefont{Gole\ifmmode~\check{z}\else
  \v{z}\fi{}}}, \bibinfo {author}
  {\bibfnamefont{J.}~\bibnamefont{Bon\ifmmode~\check{c}\else \v{c}\fi{}a}},\
  and\ \bibinfo {author} {\bibfnamefont{L.}~\bibnamefont{Vidmar}},\ }%
  \bibfield{journal}{%
  \Doi{10.1103/PhysRevB.85.144304}{\bibinfo {journal} {Phys. Rev. B}}\ }%
  \textbf{\bibinfo {volume} {85}},\ \bibinfo {pages} {144304} (\bibinfo {year}
  {2012})%
  \bibAnnoteFile{NoStop}{PhysRevB.85.144304}%
\bibitem{Vinkler2011}%
  \BibitemOpen
  \bibfield{author}{%
  \bibinfo {author} {\bibfnamefont{Y.}~\bibnamefont{Vinkler}}, \bibinfo
  {author} {\bibfnamefont{A.}~\bibnamefont{Schiller}},\ and\ \bibinfo {author}
  {\bibfnamefont{N.}~\bibnamefont{Andrei}},\ }%
  \bibfield{journal}{%
  \Doi{10.1103/PhysRevB.85.035411}{\bibinfo {journal} {Phys. Rev. B}}\ }%
  \textbf{\bibinfo {volume} {85}},\ \bibinfo {pages} {035411} (\bibinfo {year}
  {2012})%
  \bibAnnoteFile{NoStop}{Vinkler2011}%
\bibitem{JPSJ.81.013701}%
  \BibitemOpen
  \bibfield{author}{%
  \bibinfo {author} {\bibfnamefont{H.}~\bibnamefont{Matsueda}}, \bibinfo
  {author} {\bibfnamefont{S.}~\bibnamefont{Sota}}, \bibinfo {author}
  {\bibfnamefont{T.}~\bibnamefont{Tohyama}},\ and\ \bibinfo {author}
  {\bibfnamefont{S.}~\bibnamefont{Maekawa}},\ }%
  \bibfield{journal}{%
  \Doi{10.1143/JPSJ.81.013701}{\bibinfo {journal} {J. Phys. Soc. Jpn.}}\ }%
  \textbf{\bibinfo {volume} {81}},\ \bibinfo {pages} {013701} (\bibinfo {year}
  {2012})%
  \bibAnnoteFile{NoStop}{JPSJ.81.013701}%
\bibitem{PhysRevLett.109.176402}%
  \BibitemOpen
  \bibfield{author}{%
  \bibinfo {author} {\bibfnamefont{G.}~\bibnamefont{De~Filippis}}, \bibinfo
  {author} {\bibfnamefont{V.}~\bibnamefont{Cataudella}}, \bibinfo {author}
  {\bibfnamefont{E.~A.}\ \bibnamefont{Nowadnick}}, \bibinfo {author}
  {\bibfnamefont{T.~P.}\ \bibnamefont{Devereaux}}, \bibinfo {author}
  {\bibfnamefont{A.~S.}\ \bibnamefont{Mishchenko}},\ and\ \bibinfo {author}
  {\bibfnamefont{N.}~\bibnamefont{Nagaosa}},\ }%
  \bibfield{journal}{%
  \Doi{10.1103/PhysRevLett.109.176402}{\bibinfo {journal} {Phys. Rev. Lett.}}\
  }%
  \textbf{\bibinfo {volume} {109}},\ \bibinfo {pages} {176402} (\bibinfo {year}
  {2012})%
  \bibAnnoteFile{NoStop}{PhysRevLett.109.176402}%
\bibitem{PhysRevB.84.195109}%
  \BibitemOpen
  \bibfield{author}{%
  \bibinfo {author} {\bibfnamefont{J.~D.}\ \bibnamefont{Lee}}, \bibinfo
  {author} {\bibfnamefont{P.}~\bibnamefont{Moon}},\ and\ \bibinfo {author}
  {\bibfnamefont{M.}~\bibnamefont{Hase}},\ }%
  \bibfield{journal}{%
  \Doi{10.1103/PhysRevB.84.195109}{\bibinfo {journal} {Phys. Rev. B}}\ }%
  \textbf{\bibinfo {volume} {84}},\ \bibinfo {pages} {195109} (\bibinfo {year}
  {2011})%
  \bibAnnoteFile{NoStop}{PhysRevB.84.195109}%
\bibitem{PhysRevB.79.125118}%
  \BibitemOpen
  \bibfield{author}{%
  \bibinfo {author} {\bibfnamefont{K.}~\bibnamefont{Yonemitsu}}\ and\ \bibinfo
  {author} {\bibfnamefont{N.}~\bibnamefont{Maeshima}},\ }%
  \bibfield{journal}{%
  \Doi{10.1103/PhysRevB.79.125118}{\bibinfo {journal} {Phys. Rev. B}}\ }%
  \textbf{\bibinfo {volume} {79}},\ \bibinfo {pages} {125118} (\bibinfo {year}
  {2009})%
  \bibAnnoteFile{NoStop}{PhysRevB.79.125118}%
\bibitem{Schmitt2008}%
  \BibitemOpen
  \bibfield{author}{%
  \bibinfo {author} {\bibfnamefont{F.}~\bibnamefont{Schmitt}}, \bibinfo
  {author} {\bibfnamefont{P.~S.}\ \bibnamefont{Kirchmann}}, \bibinfo {author}
  {\bibfnamefont{U.}~\bibnamefont{Bovensiepen}}, \bibinfo {author}
  {\bibfnamefont{R.~G.}\ \bibnamefont{Moore}}, \bibinfo {author}
  {\bibfnamefont{L.}~\bibnamefont{Rettig}}, \bibinfo {author}
  {\bibfnamefont{M.}~\bibnamefont{Krenz}}, \bibinfo {author}
  {\bibfnamefont{J.-H.}\ \bibnamefont{Chu}}, \bibinfo {author}
  {\bibfnamefont{N.}~\bibnamefont{Ru}}, \bibinfo {author}
  {\bibfnamefont{L.}~\bibnamefont{Perfetti}}, \bibinfo {author}
  {\bibfnamefont{D.~H.}\ \bibnamefont{Lu}}, \bibinfo {author}
  {\bibfnamefont{M.}~\bibnamefont{Wolf}}, \bibinfo {author}
  {\bibfnamefont{I.~R.}\ \bibnamefont{Fisher}},\ and\ \bibinfo {author}
  {\bibfnamefont{Z.-X.}\ \bibnamefont{Shen}},\ }%
  \bibfield{journal}{%
  \Doi{10.1126/science.1160778}{\bibinfo {journal} {Science}}\ }%
  \textbf{\bibinfo {volume} {321}},\ \bibinfo {pages} {1649} (\bibinfo {year}
  {2008})%
  \bibAnnoteFile{NoStop}{Schmitt2008}%
\bibitem{Yonemitsu20081}%
  \BibitemOpen
  \bibfield{author}{%
  \bibinfo {author} {\bibfnamefont{K.}~\bibnamefont{Yonemitsu}}\ and\ \bibinfo
  {author} {\bibfnamefont{K.}~\bibnamefont{Nasu}},\ }%
  \bibfield{journal}{%
  \Doi{http://dx.doi.org/10.1016/j.physrep.2008.04.008}{\bibinfo {journal}
  {Phys. Rep.}}\ }%
  \textbf{\bibinfo {volume} {465}},\ \bibinfo {pages} {1 } (\bibinfo {year}
  {2008})%
  \bibAnnoteFile{NoStop}{Yonemitsu20081}%
\bibitem{Rubtsov05}%
  \BibitemOpen
  \bibfield{author}{%
  \bibinfo {author} {\bibfnamefont{A.~N.}\ \bibnamefont{Rubtsov}}, \bibinfo
  {author} {\bibfnamefont{V.~V.}\ \bibnamefont{Savkin}},\ and\ \bibinfo
  {author} {\bibfnamefont{A.~I.}\ \bibnamefont{Lichtenstein}},\ }%
  \bibfield{journal}{%
  \bibinfo {journal} {Phys. Rev. B}\ }%
  \textbf{\bibinfo {volume} {72}},\ \bibinfo {pages} {035122} (\bibinfo {year}
  {2005})%
  \bibAnnoteFile{NoStop}{Rubtsov05}%
\bibitem{PhysRevA.84.051401}%
  \BibitemOpen
  \bibfield{author}{%
  \bibinfo {author} {\bibfnamefont{F.}~\bibnamefont{Herrera}}\ and\ \bibinfo
  {author} {\bibfnamefont{R.~V.}\ \bibnamefont{Krems}},\ }%
  \bibfield{journal}{%
  \Doi{10.1103/PhysRevA.84.051401}{\bibinfo {journal} {Phys. Rev. A}}\ }%
  \textbf{\bibinfo {volume} {84}},\ \bibinfo {pages} {051401} (\bibinfo {year}
  {2011})%
  \bibAnnoteFile{NoStop}{PhysRevA.84.051401}%
\bibitem{1367-2630-14-3-033019}%
  \BibitemOpen
  \bibfield{author}{%
  \bibinfo {author} {\bibfnamefont{J.~P.}\ \bibnamefont{Hague}}\ and\ \bibinfo
  {author} {\bibfnamefont{C.}~\bibnamefont{MacCormick}},\ }%
  \bibfield{journal}{%
  \bibinfo {journal} {New Journal of Physics}\ }%
  \textbf{\bibinfo {volume} {14}},\ \bibinfo {pages} {033019} (\bibinfo {year}
  {2012})%
  \bibAnnoteFile{NoStop}{1367-2630-14-3-033019}%
\bibitem{PhysRevA.72.033609}%
  \BibitemOpen
  \bibfield{author}{%
  \bibinfo {author} {\bibfnamefont{E.}~\bibnamefont{Pazy}}\ and\ \bibinfo
  {author} {\bibfnamefont{A.}~\bibnamefont{Vardi}},\ }%
  \bibfield{journal}{%
  \Doi{10.1103/PhysRevA.72.033609}{\bibinfo {journal} {Phys. Rev. A}}\ }%
  \textbf{\bibinfo {volume} {72}},\ \bibinfo {pages} {033609} (\bibinfo {year}
  {2005})%
  \bibAnnoteFile{NoStop}{PhysRevA.72.033609}%
\bibitem{BlochSSH}%
  \BibitemOpen
  \bibfield{author}{%
  \bibinfo {author} {\bibfnamefont{M.}~\bibnamefont{Atala}}, \bibinfo {author}
  {\bibfnamefont{M.}~\bibnamefont{Aidelsburger}}, \bibinfo {author}
  {\bibfnamefont{J.~T.}\ \bibnamefont{Barreiro}}, \bibinfo {author}
  {\bibfnamefont{D.}~\bibnamefont{Abanin}}, \bibinfo {author}
  {\bibfnamefont{T.}~\bibnamefont{Kitagawa}}, \bibinfo {author}
  {\bibfnamefont{E.}~\bibnamefont{Demler}},\ and\ \bibinfo {author}
  {\bibfnamefont{I.}~\bibnamefont{Bloch}},\ }%
  \bibfield{journal}{%
  \bibinfo {journal} {arXiv:1212.0572}}%
  \bibAnnoteFile{NoStop}{BlochSSH}%
\bibitem{PhysRevLett.109.200501}%
  \BibitemOpen
  \bibfield{author}{%
  \bibinfo {author} {\bibfnamefont{A.}~\bibnamefont{Mezzacapo}}, \bibinfo
  {author} {\bibfnamefont{J.}~\bibnamefont{Casanova}}, \bibinfo {author}
  {\bibfnamefont{L.}~\bibnamefont{Lamata}},\ and\ \bibinfo {author}
  {\bibfnamefont{E.}~\bibnamefont{Solano}},\ }%
  \bibfield{journal}{%
  \Doi{10.1103/PhysRevLett.109.200501}{\bibinfo {journal} {Phys. Rev. Lett.}}\
  }%
  \textbf{\bibinfo {volume} {109}},\ \bibinfo {pages} {200501} (\bibinfo {year}
  {2012})%
  \bibAnnoteFile{NoStop}{PhysRevLett.109.200501}%
\bibitem{PhysRevLett.109.250501}%
  \BibitemOpen
  \bibfield{author}{%
  \bibinfo {author} {\bibfnamefont{V.~M.}\
  \bibnamefont{Stojanovi\ifmmode~\acute{c}\else \'{c}\fi{}}}, \bibinfo {author}
  {\bibfnamefont{T.}~\bibnamefont{Shi}}, \bibinfo {author}
  {\bibfnamefont{C.}~\bibnamefont{Bruder}},\ and\ \bibinfo {author}
  {\bibfnamefont{J.~I.}\ \bibnamefont{Cirac}},\ }%
  \bibfield{journal}{%
  \Doi{10.1103/PhysRevLett.109.250501}{\bibinfo {journal} {Phys. Rev. Lett.}}\
  }%
  \textbf{\bibinfo {volume} {109}},\ \bibinfo {pages} {250501} (\bibinfo {year}
  {2012})%
  \bibAnnoteFile{NoStop}{PhysRevLett.109.250501}%
\bibitem{Ho59a}%
  \BibitemOpen
  \bibfield{author}{%
  \bibinfo {author} {\bibfnamefont{T.}~\bibnamefont{Holstein}},\ }%
  \bibfield{journal}{%
  \bibinfo {journal} {Ann. Phys. (N.Y.)}\ }%
  \textbf{\bibinfo {volume} {8}},\ \bibinfo {pages} {325; {\bf 8}, 343}
  (\bibinfo {year} {1959})%
  \bibAnnoteFile{NoStop}{Ho59a}%
\bibitem{Gull_rev}%
  \BibitemOpen
  \bibfield{author}{%
  \bibinfo {author} {\bibfnamefont{E.}~\bibnamefont{Gull}}, \bibinfo {author}
  {\bibfnamefont{A.~J.}\ \bibnamefont{Millis}}, \bibinfo {author}
  {\bibfnamefont{A.~I.}\ \bibnamefont{Lichtenstein}}, \bibinfo {author}
  {\bibfnamefont{A.~N.}\ \bibnamefont{Rubtsov}}, \bibinfo {author}
  {\bibfnamefont{M.}~\bibnamefont{Troyer}},\ and\ \bibinfo {author}
  {\bibfnamefont{P.}~\bibnamefont{Werner}},\ }%
  \bibfield{journal}{%
  \Doi{10.1103/RevModPhys.83.349}{\bibinfo {journal} {Rev. Mod. Phys.}}\ }%
  \textbf{\bibinfo {volume} {83}},\ \bibinfo {pages} {349} (\bibinfo {year}
  {2011})%
  \bibAnnoteFile{NoStop}{Gull_rev}%
\bibitem{Assaad07}%
  \BibitemOpen
  \bibfield{author}{%
  \bibinfo {author} {\bibfnamefont{F.~F.}\ \bibnamefont{Assaad}}\ and\ \bibinfo
  {author} {\bibfnamefont{T.~C.}\ \bibnamefont{Lang}},\ }%
  \bibfield{journal}{%
  \Doi{10.1103/PhysRevB.76.035116}{\bibinfo {journal} {Phys. Rev. B}}\ }%
  \textbf{\bibinfo {volume} {76}},\ \bibinfo {eid} {035116} (\bibinfo {year}
  {2007})%
  \bibAnnoteFile{NoStop}{Assaad07}%
\bibitem{Hohenadler10a}%
  \BibitemOpen
  \bibfield{author}{%
  \bibinfo {author} {\bibfnamefont{M.}~\bibnamefont{Hohenadler}}, \bibinfo
  {author} {\bibfnamefont{H.}~\bibnamefont{Fehske}},\ and\ \bibinfo {author}
  {\bibfnamefont{F.~F.}\ \bibnamefont{Assaad}},\ }%
  \bibfield{journal}{%
  \Doi{10.1103/PhysRevB.83.115105}{\bibinfo {journal} {Phys. Rev. B}}\ }%
  \textbf{\bibinfo {volume} {83}},\ \bibinfo {pages} {115105} (\bibinfo {year}
  {2011})%
  \bibAnnoteFile{NoStop}{Hohenadler10a}%
\bibitem{Assaad08}%
  \BibitemOpen
  \bibfield{author}{%
  \bibinfo {author} {\bibfnamefont{F.~F.}\ \bibnamefont{Assaad}},\ }%
  \bibfield{journal}{%
  \Doi{10.1103/PhysRevB.78.155124}{\bibinfo {journal} {Phys. Rev. B}}\ }%
  \textbf{\bibinfo {volume} {78}},\ \bibinfo {eid} {155124} (\bibinfo {year}
  {2008})%
  \bibAnnoteFile{NoStop}{Assaad08}%
\bibitem{HoAs12}%
  \BibitemOpen
  \bibfield{author}{%
  \bibinfo {author} {\bibfnamefont{M.}~\bibnamefont{Hohenadler}}\ and\ \bibinfo
  {author} {\bibfnamefont{F.~F.}\ \bibnamefont{Assaad}},\ }%
  \bibfield{journal}{%
  \bibinfo {journal} {J. Phys.: Condens. Matter}\ }%
  \textbf{\bibinfo {volume} {25}},\ \bibinfo {pages} {014005} (\bibinfo {year}
  {2013})%
  \bibAnnoteFile{NoStop}{HoAs12}%
\bibitem{PhysRevB.87.075149}%
  \BibitemOpen
  \bibfield{author}{%
  \bibinfo {author} {\bibfnamefont{M.}~\bibnamefont{Hohenadler}}\ and\ \bibinfo
  {author} {\bibfnamefont{F.~F.}\ \bibnamefont{Assaad}},\ }%
  \bibfield{journal}{%
  \Doi{10.1103/PhysRevB.87.075149}{\bibinfo {journal} {Phys. Rev. B}}\ }%
  \textbf{\bibinfo {volume} {87}},\ \bibinfo {pages} {075149} (\bibinfo {year}
  {2013})%
  \bibAnnoteFile{NoStop}{PhysRevB.87.075149}%
\bibitem{Ho.As.Fe.12}%
  \BibitemOpen
  \bibfield{author}{%
  \bibinfo {author} {\bibfnamefont{M.}~\bibnamefont{Hohenadler}}, \bibinfo
  {author} {\bibfnamefont{F.~F.}\ \bibnamefont{Assaad}},\ and\ \bibinfo
  {author} {\bibfnamefont{H.}~\bibnamefont{Fehske}},\ }%
  \bibfield{journal}{%
  \bibinfo {journal} {Phys. Rev. Lett.}\ }%
  \textbf{\bibinfo {volume} {109}},\ \bibinfo {pages} {116407} (\bibinfo {year}
  {2012})%
  \bibAnnoteFile{NoStop}{Ho.As.Fe.12}%
\bibitem{PhysRevB.85.085129}%
  \BibitemOpen
  \bibfield{author}{%
  \bibinfo {author} {\bibfnamefont{F.}~\bibnamefont{Goth}}\ and\ \bibinfo
  {author} {\bibfnamefont{F.~F.}\ \bibnamefont{Assaad}},\ }%
  \bibfield{journal}{%
  \Doi{10.1103/PhysRevB.85.085129}{\bibinfo {journal} {Phys. Rev. B}}\ }%
  \textbf{\bibinfo {volume} {85}},\ \bibinfo {pages} {085129} (\bibinfo {year}
  {2012})%
  \bibAnnoteFile{NoStop}{PhysRevB.85.085129}%
\bibitem{PhysRevLett.100.176403}%
  \BibitemOpen
  \bibfield{author}{%
  \bibinfo {author} {\bibfnamefont{L.}~\bibnamefont{M\"uhlbacher}}\ and\
  \bibinfo {author} {\bibfnamefont{E.}~\bibnamefont{Rabani}},\ }%
  \bibfield{journal}{%
  \Doi{10.1103/PhysRevLett.100.176403}{\bibinfo {journal} {Phys. Rev. Lett.}}\
  }%
  \textbf{\bibinfo {volume} {100}},\ \bibinfo {pages} {176403} (\bibinfo {year}
  {2008})%
  \bibAnnoteFile{NoStop}{PhysRevLett.100.176403}%
\bibitem{Hirsch83a}%
  \BibitemOpen
  \bibfield{author}{%
  \bibinfo {author} {\bibfnamefont{J.~E.}\ \bibnamefont{Hirsch}}\ and\ \bibinfo
  {author} {\bibfnamefont{E.}~\bibnamefont{Fradkin}},\ }%
  \bibfield{journal}{%
  \bibinfo {journal} {Phys. Rev. B}\ }%
  \textbf{\bibinfo {volume} {27}},\ \bibinfo {pages} {4302} (\bibinfo {year}
  {1983})%
  \bibAnnoteFile{NoStop}{Hirsch83a}%
\bibitem{JeZhWh99}%
  \BibitemOpen
  \bibfield{author}{%
  \bibinfo {author} {\bibfnamefont{E.}~\bibnamefont{Jeckelmann}}, \bibinfo
  {author} {\bibfnamefont{C.}~\bibnamefont{Zhang}},\ and\ \bibinfo {author}
  {\bibfnamefont{S.~R.}\ \bibnamefont{White}},\ }%
  \bibfield{journal}{%
  \bibinfo {journal} {Phys. Rev. B}\ }%
  \textbf{\bibinfo {volume} {60}},\ \bibinfo {pages} {7950} (\bibinfo {year}
  {1999})%
  \bibAnnoteFile{NoStop}{JeZhWh99}%
\bibitem{0295-5075-84-5-57001}%
  \BibitemOpen
  \bibfield{author}{%
  \bibinfo {author} {\bibfnamefont{H.}~\bibnamefont{Fehske}}, \bibinfo {author}
  {\bibfnamefont{G.}~\bibnamefont{Hager}},\ and\ \bibinfo {author}
  {\bibfnamefont{E.}~\bibnamefont{Jeckelmann}},\ }%
  \bibfield{journal}{%
  \bibinfo {journal} {Europhys. Lett.}\ }%
  \textbf{\bibinfo {volume} {84}},\ \bibinfo {pages} {57001} (\bibinfo {year}
  {2008})%
  \bibAnnoteFile{NoStop}{0295-5075-84-5-57001}%
\bibitem{ClHa05}%
  \BibitemOpen
  \bibfield{author}{%
  \bibinfo {author} {\bibfnamefont{R.~T.}\ \bibnamefont{Clay}}\ and\ \bibinfo
  {author} {\bibfnamefont{R.~P.}\ \bibnamefont{Hardikar}},\ }%
  \bibfield{journal}{%
  \bibinfo {journal} {Phys. Rev. Lett.}\ }%
  \textbf{\bibinfo {volume} {95}},\ \bibinfo {pages} {096401} (\bibinfo {year}
  {2005})%
  \bibAnnoteFile{NoStop}{ClHa05}%
\bibitem{hardikar:245103}%
  \BibitemOpen
  \bibfield{author}{%
  \bibinfo {author} {\bibfnamefont{R.~P.}\ \bibnamefont{Hardikar}}\ and\
  \bibinfo {author} {\bibfnamefont{R.~T.}\ \bibnamefont{Clay}},\ }%
  \bibfield{journal}{%
  \Doi{10.1103/PhysRevB.75.245103}{\bibinfo {journal} {Phys. Rev. B}}\ }%
  \textbf{\bibinfo {volume} {75}},\ \bibinfo {eid} {245103} (\bibinfo {year}
  {2007})%
  \bibAnnoteFile{NoStop}{hardikar:245103}%
\bibitem{Schulz90}%
  \BibitemOpen
  \bibfield{author}{%
  \bibinfo {author} {\bibfnamefont{H.~J.}\ \bibnamefont{Schulz}},\ }%
  \bibfield{journal}{%
  \bibinfo {journal} {Phys. Rev. Lett}\ }%
  \textbf{\bibinfo {volume} {64}},\ \bibinfo {pages} {2831} (\bibinfo {year}
  {1990})%
  \bibAnnoteFile{NoStop}{Schulz90}%
\bibitem{PhysRevLett.98.050405}%
  \BibitemOpen
  \bibfield{author}{%
  \bibinfo {author} {\bibfnamefont{M.}~\bibnamefont{Rigol}}, \bibinfo {author}
  {\bibfnamefont{V.}~\bibnamefont{Dunjko}}, \bibinfo {author}
  {\bibfnamefont{V.}~\bibnamefont{Yurovsky}},\ and\ \bibinfo {author}
  {\bibfnamefont{M.}~\bibnamefont{Olshanii}},\ }%
  \bibfield{journal}{%
  \Doi{10.1103/PhysRevLett.98.050405}{\bibinfo {journal} {Phys. Rev. Lett.}}\
  }%
  \textbf{\bibinfo {volume} {98}},\ \bibinfo {pages} {050405} (\bibinfo {year}
  {2007})%
  \bibAnnoteFile{NoStop}{PhysRevLett.98.050405}%
\bibitem{EnnsSirker2012}%
  \BibitemOpen
  \bibfield{author}{%
  \bibinfo {author} {\bibfnamefont{T.}~\bibnamefont{Enss}}\ and\ \bibinfo
  {author} {\bibfnamefont{J.}~\bibnamefont{Sirker}},\ }%
  \bibfield{journal}{%
  \bibinfo {journal} {New J. Phys.}\ }%
  \textbf{\bibinfo {volume} {14}},\ \bibinfo {pages} {023008} (\bibinfo {year}
  {2012})%
  \bibAnnoteFile{NoStop}{EnnsSirker2012}%
\bibitem{PhysRevA.85.032114}%
  \BibitemOpen
  \bibfield{author}{%
  \bibinfo {author} {\bibfnamefont{J.}~\bibnamefont{H\"app\"ol\"a}}, \bibinfo
  {author} {\bibfnamefont{G.~B.}\ \bibnamefont{Hal\'asz}},\ and\ \bibinfo
  {author} {\bibfnamefont{A.}~\bibnamefont{Hamma}},\ }%
  \bibfield{journal}{%
  \Doi{10.1103/PhysRevA.85.032114}{\bibinfo {journal} {Phys. Rev. A}}\ }%
  \textbf{\bibinfo {volume} {85}},\ \bibinfo {pages} {032114} (\bibinfo {year}
  {2012})%
  \bibAnnoteFile{NoStop}{PhysRevA.85.032114}%
\bibitem{LiebRobinson72}%
  \BibitemOpen
  \bibfield{author}{%
  \bibinfo {author} {\bibfnamefont{E.}~\bibnamefont{Lieb}}\ and\ \bibinfo
  {author} {\bibfnamefont{D.}~\bibnamefont{Robinson}},\ }%
  \bibfield{journal}{%
  \Doi{10.1007/BF01645779}{\bibinfo {journal} {Comm. Math. Phys.}}\ }%
  \textbf{\bibinfo {volume} {28}},\ \bibinfo {pages} {251} (\bibinfo {year}
  {1972})%
  \bibAnnoteFile{NoStop}{LiebRobinson72}%
\bibitem{Giamarchi}%
  \BibitemOpen
  \bibfield{author}{%
  \bibinfo {author} {\bibfnamefont{T.}~\bibnamefont{Giamarchi}},\ }%
  \emph{\bibinfo {title} {Quantum Physics in One Dimension}}\ (\bibinfo
  {publisher} {Clarendon Press},\ \bibinfo {address} {Oxford},\ \bibinfo {year}
  {2004})%
  \bibAnnoteFile{NoStop}{Giamarchi}%
\end{thebibliography}
%

\end{document}